# High performance Microreactor for Rapid Fluid Mixing and Redox Reaction of Ascorbic Acid

W. F. Fang[1] and J. T. Yang[1, 2]
[1]Department of Power Mechanical Engineering, National Tsing Hua University, Hsinchu 30013, Taiwan
[2]Institute of NanoEngineering and MicroSystems, National Tsing Hua University, Hsinchu 30013, Taiwan
d933706@oz.nthu.edu.tw, Tel: +886-3-5715131, Fax: +886-3-5724242.

*Abstract*- **A novel micro device with a mechanism of split and recombination (SNR) for rapid fluidic mixing and reaction, named a SNR µ-reactor, was designed, fabricated and systematically analyzed. This SNR µ-reactor possessing an in-plane dividing structure requires only simple fabrication. We investigated this reactor and compared it numerically and experimentally with a slanted-groove micromixer (SGM). From the numerical results the mixing indices and mixing patterns demonstrated that the mixing ability of the SNR µ-reactor was much superior to that of the SGM. From a mixing test with phenolphthalein and sodium hydroxide solutions, the mixing lengths of the SNR µ-reactor were less than 4 mm for a Reynolds number over a wide range (*Re* = 0.1 – 10). From a comparison of mixing lengths, the results revealed also that the SNR µ-reactor surpassed the SGM in mixing performance by more than 200 %. As a reaction length is a suitable test of the performance of a reactor, we introduced a redox reaction between ascorbic acid and iodine solutions to assess the reaction capability of these micro devices; the reaction lengths of the SNR µ-reactor were much shorter than those of a SGM. The SNR µ-reactor has consequently a remarkable efficiency for fluid mixing and reaction.**

*Presentation*—Oral

## I. Introduction

Because its ratio of surface to volume is large, a microfluidic system possesses many applications and advantages of small sample consumption, parallel handling, and precise and rapid detection. In the treatment of biological or chemical samples, mixing is a crucial procedure. Rapid and uniform mixing of samples not only expedites reaction but also promotes the accuracy of assay.

To improve mixing in a micro-device the major mechanism involves enlarging the contact interface, thus decreasing the mixing length and period. Named micromixers or microreactors, these devices are classified as operating actively or passively [1]. Passive micromixers rely mainly on the principles of chaotic advection [2, 3], split and recombination (SNR) [4, 5], and fluidic instability [6]. The factors that affect chemical or biological reactions are the concentrations of reactants, the intrinsic properties, the operating temperature, the catalyst and the contact area; increasing the contact area can especially bolster reactions.

We hence proposed a novel microreactor with a mixing mechanism involving split and recombination. Called a SNR µ-reactor, this microreactor enables rapid mixing and the reaction of fluids for Reynolds number over a wide range. We not only employed numerical simulation to analyze these devices but also demonstrated quantitatively their performance via chemical experiments – specifically, with an acid-base indicator and a redox reaction of ascorbic acid.

## II. Design Concept, Fabrication and Experiments

The design concept and mixing mechanisms of a SNR µ-reactor are shown in Fig. 1. The upper (Fluid 1) and lower (Fluid 2) fluids overlap and combine at a confluence according to process *A*. The fluids then flow downstream to be divided by the in-plane dividing edge as process *B*; the divided fluids are subsequently guided apart along separate channels, and they eventually recombine at the next confluence as process *C*. For the numerical simulation, a SNR µ-reactor comprised ten mixing units and had a total length 6100 µm (Fig. 2a). Figs. 2(b) and (c) present the detailed dimensions of a mixing unit.

Fig. 3(a) shows a fabrication flowchart of a SNR µ-reactor. According to standard fabrication procedures with a thick photo-resist (SU-8 2035, negative photo-resist), a complementary SU-8 mold of the SNR µ-reactor was fabricated via single-step photolithography. We poured the degassed polydimethylsiloxane (PDMS) mixture into the mold and then peeled it off after solidification. After treating the PDMS with an oxygen plasma, the bonding of the device was easily performed. Fig. 3(b) exhibits SEM photographs of the channel structure and the detailed dimensions. We utilized commercial software (CFD-ACE+, CFD Research Corp.) to simulate the fluidic system. All relevant numerical setup and analysis of sensitivity to the grid size were conducted based on the results of Yang et al. (2005) [7].

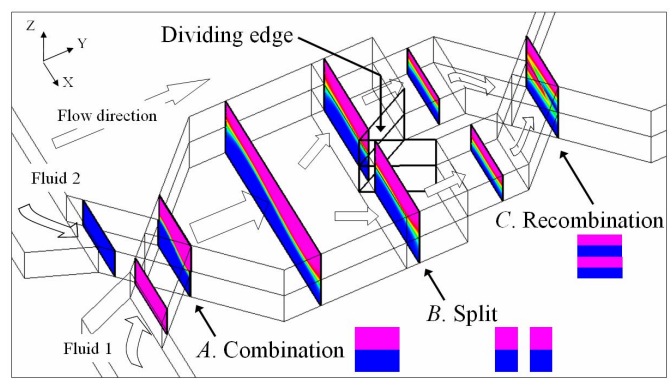

Fig. 1 Schematic diagram of the design concept for a SNR µ-reactor.

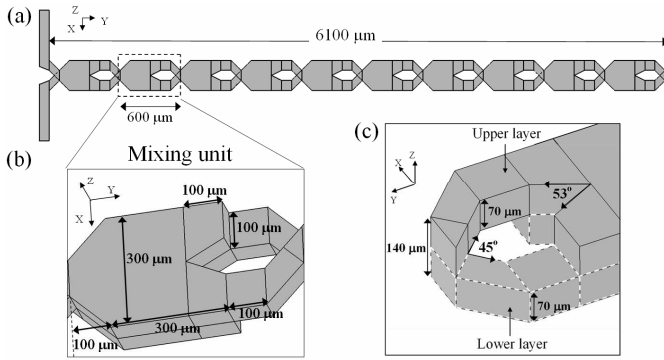

Fig. 2 Configuration of a SNR μ-reactor for numerical simulation. (a) Overview of SNR μ-reactor; (b) (c) dimensions of a mixing unit.

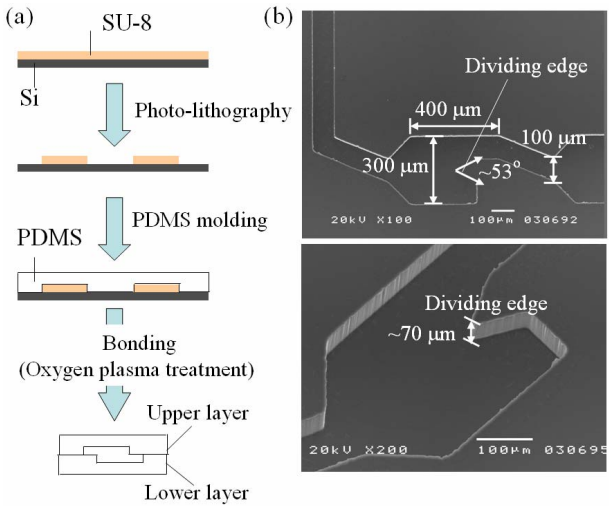

Fig. 3 (a) Fabrication flowchart of devices; (b) SEM photos of the microchannel of a SNR μ-reactor.

## III. RESULTS AND DISCUSSION

### A. Analysis of mixing performance from simulation

A mixing index (Mi) is generally used to represent quantitatively the ability of a device to mix fluids. A mixing index 0.9 is accepted to indicate effective mixing of fluids [2]. Fig. 4 shows the mixing indices of a SNR μ-reactor and a slanted-groove micromixer (SGM) at $Re = 1$ with their mixing patterns at various $Y$ positions ($Y = 0, 2, 6$ and $10$ mm). The SNR μ-reactor required a length about 4 mm to achieve effective mixing of fluids, whereas the SGM required a length over 10 mm to achieve satisfactory mixing. The mixing performance of the SNR μ-reactor is clearly much better than that of the SGM.

According to the mixing patterns of fluids, mass transfer of fluids relies mainly on the structural design of the devices. Modifying the mixing patterns along the downstream channel assists the continuing mixing of fluids within devices. In the SGM, the slanted grooves induce mass transfer of fluids only at the bottom of the channel; the evolution of mixing patterns is consequently so weak that mixing efficiency is unsatisfactory. As for the SNR μ-reactor, its variation of mixing patterns was intense because of the mechanism of the SNR reflecting its overlapping channel and dividing structures. The patterns show that the SNR μ-reactor possesses a remarkable mixing efficiency for microfluids.

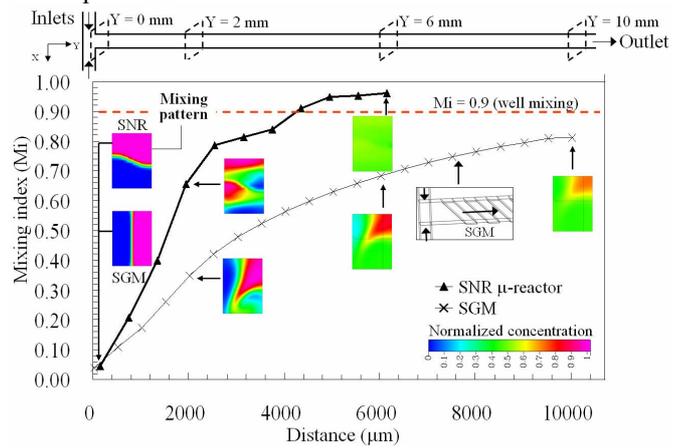

Fig. 4 Mixing indices and mixing patterns of a SNR μ-reactor and a SGM at $Re = 1$.

### B. Verification of mixing and reaction efficiency from experiments

A rapid chemical reaction involving phenolphthalein solution is commonly used to investigate the mixing performance of micromixers [8]. The slanted grooves at the bottom of the channel induce a helical flow so that one can perceive spiral reacting interfaces in SGM as shown in Fig. 5(a). The mixing of fluids in a SGM was evidently incomplete at $Y \sim 4$ mm whereas fluids were fully mixed in a SNR μ-reactor as shown in Fig. 5 (b). A mixing length is customarily determined based on the distance of variation of gray-level intensity downstream via image processing [8]. We characterized the relation between the mixing length and the Reynolds number as shown in Fig. 6. The mixing lengths of a SNR μ-reactor were less than 4 mm and much shorter than that of a SGM for $Re$ from 0.1 to 10. The difference in mixing lengths between the SNR and the SGM increased with increasing Reynolds number. The SNR μ-reactor is clearly highly efficient and applicable to mix fluids with $Re$ in a wide range.

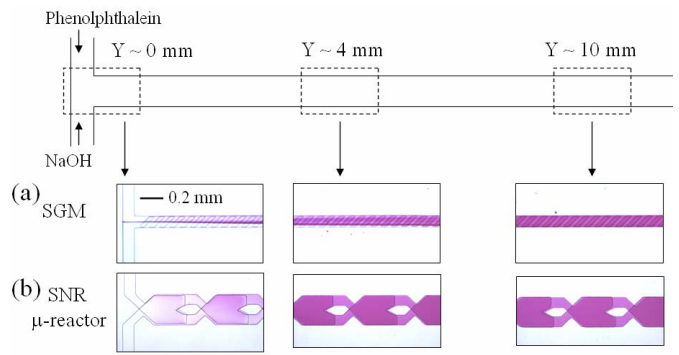

Fig. 5 Experimental results of (a) a SNR μ-reactor; (b) a SGM, based on mixing of phenolphthalein (~ 0.3 M) and sodium hydroxide solution (~ 0.3 M, pH = 13) at a flow rate 10 μL/min ($Re \sim 1.6$).

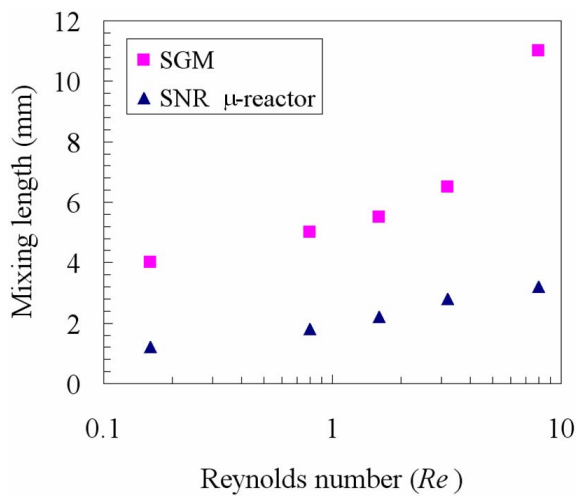

Fig. 6 Mixing lengths of SNR µ-reactor and SGM at various $Re$ based on mixing experiments of phenolphthalein.

To assess the reaction performance of devices, we introduced a novel approach involving a redox reaction of ascorbic acid. We adopted diiodine as an oxidizing agent for ascorbic acid whereby diiodine becomes reduced by ascorbic acid in solution; the hues of the solutions alter concurrently from orange to colorless as shown in Fig. 7. The preparation of the iodine solution is based on equation (1) and the nature of the reaction is illustrated in equation (2). We defined a distance of color change from orange to colorless as a reaction length. A reaction period is derived from a mean velocity divided by the reaction length. As shown in Fig. 7(a) lower, solutions of diiodine (0.35 M) and ascorbic acid (0.5 M) underwent six mixing units in a SNR µ-reactor to achieve complete reaction (reaction length about 3.6 mm) at a flow rate 10 µL/min; in contrast, the reaction length of SGM was more than 12 mm (see Fig. 7(b)). The reaction period of the redox process in a SNR µ-reactor was hence less than that in a SGM. When the concentration of ascorbic acid was increased to 1.0 M, the reaction length of the SNR µ-reactor decreased from 3.6 mm to 2.4 mm as shown in Fig. 7(c); the reaction length of the SGM was decreased to about 9 mm (see Fig. 7(d)). The concentration of the reductant thus crucially influences a redox reaction. The results confirm also that the SNR µ-reactor produced an efficiency of fluid mixing and reaction superior to that of the SGM.

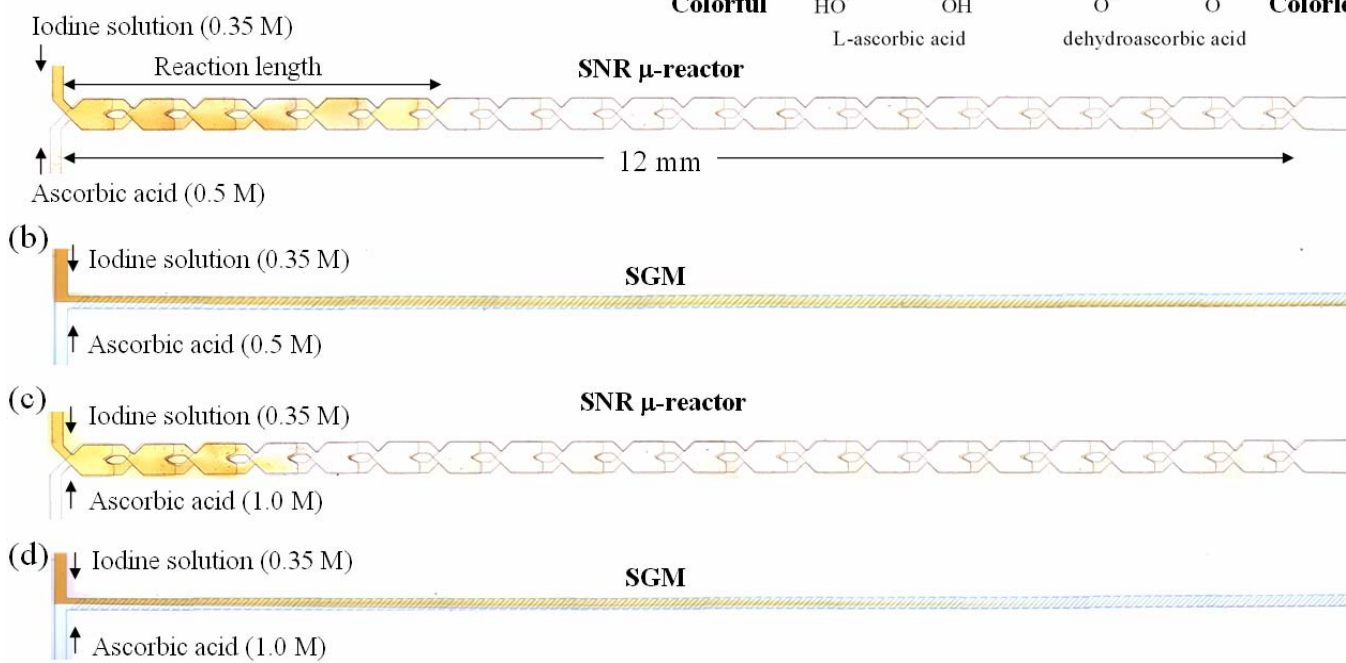

Fig. 7 Experiments for redox reactions of ascorbic acid and iodine solutions. The chemical equations (1) and (2) represent the preparation of an iodine solution and the redox reaction, respectively. Reaction results for an iodine solution, 0.35 M and ascorbic acid, 0.5 M in (a) SNR µ-reactor; (b) SGM. Reaction results for an iodine solution, 0.35 M, and ascorbic acid, 1.0 M, in (c) SNR µ-reactor; (d) SGM at a flow rate 10 µL/min.

## IV. CONCLUSIONS

Our SNR µ-reactor demonstrated a mixing and reaction performance superior to that of a slanted-groove micromixer (SGM) for a Reynolds number over a wide range; moreover, its fabrication was simple. The simulation results illustrated that intense mixing of fluids occurred within the SNR µ-reactor; the lengths of effective mixing for the SNR µ-reactor were consequently much smaller than that for the SGM. According to a comparison of mixing lengths based on a test of phenolphthalein solution, the mixing ability of SNR μ-reactor was at least twice as good as that of the SGM for $Re = 0.1 \sim 10$. The mixing lengths of the SNR μ-reactor were shorter than 4 mm. A redox reaction of ascorbic acid with diiodine solution was introduced to assess the reaction efficiency of micro devices through estimation of the reaction length. The results demonstrate explicitly that the SNR μ-reactor produced satisfactory reaction lengths and reaction durations.

In sum, this SNR μ-reactor has perfect mixing and reaction efficiency; it can consequently be applied for the rapid reaction of chemical and biological fluids. This novel microreactor has a great potential for applications in a microfluidic system. We are now endeavoring to exploit the SNR μ-reactor to enhance the hybridization efficiency of DNA for rapid and sensitive detection in a micro system.

ACKNOWLEDGMENT

National Science Council of the Republic of China supported this work under contracts NSC 96-2628-E007-120-MY3 and NSC 96-2628-E007-121-MY3.

BIOGRAPHY

Born in Taiwan, R.O.C., in 1981, Wei-Feng Fang was awarded the B.S. degree by National Chung Cheng University, Min-Hsiung Chia-Yi, Taiwan, in 2004. He is at present working towards the Ph. D. degree in the Department of Power Mechanical Engineering of National Tsing Hua University, Hsinchu, Taiwan. His research interests include design of microfluidic systems, microfluid mixing and reaction systems, and bioMEMS